\documentclass[12pt]{article}

\usepackage{amsmath}
\usepackage{amsthm}
\usepackage{amssymb}
\usepackage{amsfonts}
\usepackage[latin1]{inputenc}

\def\be{\begin{equation}}
\def\ee{\end{equation}}
\def\bea{\begin{eqnarray}}
\def\eea{\end{eqnarray}}

\newcommand{\sect}[1]{\setcounter{equation}{0}\section{#1}}

\newcommand{\rsl}{\mathfrak{sl}}
 
\newcommand{\bq}{\mathbf{q}}
\newcommand{\bp}{\mathbf{p}}

\def\RR{\mathbb{R}}
\newcommand{\cH}{{\mathcal H}}
\newcommand{\cM}{{\mathcal M}}
\def\dd{{\rm d}}
\newcommand{\te}{\theta}
\newcommand\Om\Omega
\newcommand{\bL}{\mathbf{L}}
\newcommand{\De}{\Delta}
\newcommand\dif[2]{\frac{\pd #1}{\pd #2}}
\newcommand{\pd}{\partial}
\newcommand\minus\backslash

\newcommand{\Kep}{_{\mathrm{KC}}}
\newcommand{\Harm}{_{\mathrm{O}}}
\newcommand{\sub}[1]{_{\mathrm{#1}}}

\def\cte{\alpha}
\def\cteb{\beta}
\def\ga{\gamma}

\def\1{\'{\i}}

\def\k{{\kappa}}
\def\kk{{k}}
\def\la{{\mu^2}}
\def\>#1{{\mathbf#1}}

\begin{document}

 \

\noindent
{\Large{\sc{Superintegrability on $N$-dimensional curved spaces: \\[6pt]
Central potentials, centrifugal terms and monopoles}}}

\bigskip
\bigskip

\begin{center}
{\'Angel Ballesteros$^a$, Alberto Enciso$^b$, Francisco J. Herranz$^c$
and Orlando Ragnisco$^d$}

\medskip

\small
\noindent
{$^a$ Depto.~de Física, Facultad de Ciencias, Universidad de Burgos,
09001 Burgos, Spain\\ ~~E-mail: angelb@ubu.es\\
$^b$ Depto.~de F{í}sica Teórica II,   Universidad Complutense,   28040 Madrid,
Spain\\ ~~E-mail: aenciso@fis.ucm.es\\
$^c$ Depto.~de Física,  Escuela Politécnica Superior, Universidad de Burgos,
09001 Burgos, Spain \\ ~~E-mail: fjherranz@ubu.es\\
$^d$ Dipartimento di Fisica,   Università di Roma Tre and Istituto Nazionale di
Fisica Nucleare sezione di Roma Tre,  Via Vasca Navale 84,  00146 Roma, Italy  \\
~~E-mail: ragnisco@fis.uniroma3.it}
\end{center}

\normalsize

\medskip

\begin{abstract}
\noindent
The $N$-dimensional Hamiltonian
$$
{\cal H}=\frac{1}{2f(|\bq|)^2}\left\{
\bp^2+ \frac{\la}{\bq^2}+\sum_{i=1}^N \frac{b_i}{q_i^2}\right\}+{\cal U}(|\bq|)
$$
is shown to be quasi-maximally superintegrable for any choice of the functions $f$ and $\cal U$. This result is proven by making use of the underlying $\mathfrak{sl}(2,\RR)$-coalgebra symmetry of ${\cal H}$ in order to obtain a set of (2$N-3$)
functionally independent integrals of the motion, that are explicitly given. Such constants of the motion are ``universal" since all of them are independent of both $f$ and $\cal U$. 
This Hamiltonian describes the motion of a particle on any $N$D spherically symmetric curved space (whose metric is specified by $f$) under the action of an arbitrary central potental $\cal U$, and includes simultaneously a monopole-type contribution together with  $N$ centrifugal   terms that break the spherical symmetry.     Moreover, we show that two appropriate choices for $\cal U$ provide the ``intrinsic"  oscillator and the KC potentials on these curved manifolds. As a byproduct,   the  MIC--Kepler, the Taub-NUT  and the so called multifold Kepler systems are shown to belong to this class of superintegrable Hamiltonians,  and  new generalizations thereof are obtained. The KC and oscillator potentials on $N$-dimensional generalizations of the four Darboux surfaces are discussed as well. 
 \end{abstract}

\bigskip   



\newpage



\section{Introduction}

An $N$-dimensional ($N$D) Hamiltonian $\cal H$ is said to be {\em (Liouville)
integrable}~\cite{AKN97} if it admits the maximum number $(N-1)$ of functionally
independent and Poisson-commuting global first integrals. Similarly, an integrable Hamiltonian $\cal H$ is called
{\em superintegrable} if there exists an additional set of $k$ 
independent constants of motion. It is well-known that $k\leq
N-1$, and in the case where this bound is saturated ($k=N-1$), the Hamiltonian $\cal H$ is called     {\em maximally superintegrable} (hereafter MS). When $k=N-2$, we shall say that $\cal H$ is   {\em quasi-maximally superintegrable} (QMS)~\cite{BH07, sigmaorlando} and the Hamiltonian $\cal H$ is only ``one integral away" from being MS.

Superintegrable systems  have
been thoroughly studied because of their significant connections with
generalized symmetries~\cite{GLZ92,STW01}, isochronous
potentials~\cite{Ca04,Go04}, and separability of the associated
Hamilton--Jacobi  and Schrödinger
equations~\cite{Hu76,KM84,BMS01}.  When one considers a Hamiltonian system in arbitrary dimension $N$ and {\em all} the integrals of motion are imposed to be {\em quadratic} in the momenta, the list of known $N$D MS systems becomes
strikingly short. To the best of our knowledge, the only known instances are: 

\begin{itemize}
\item The {\em free motion} on the simply connected {\em spaces of constant curvature}~(see, e.g.,~\cite{BHSS03,vulpiPAN} for a unified approach in terms of the curvature) which is given by the geodesic flow on these spaces.
  
\item  The
{\em generalized Kepler--Coulomb (KC) system} on the simply connected {\em spaces of constant curvature}~\cite{vulpiPAN,Ev90,Miguel, Williamsx, kiev}, which is a superposition of the $N$D KC potential with $N-1$ ``centrifugal" terms.

\item  The {\em Smorodinsky--Winternitz  system} on the simply connected {\em spaces of constant curvature}~\cite{BHSS03, vulpiPAN, kiev,FMSUW65, Ev90b, 8, 10, 11, 20, RS, 21,18}, that is,   the $N$D harmonic oscillator potential on such spaces together with $N$ centrifugal terms.

\item The {\em free motion} and the {\em Smorodinsky--Winternitz system on the Darboux space  III}~\cite{PhysD}, which is the superposition of an ``intrinsic"   oscillator and $N$ centrifugal potentials on an $N$D generalization of the Darboux surface of type III, which is a Riemannian manifold of {\em nonconstant} curvature.

\end{itemize}
Note that the above MS systems   on   spaces of {\em constant curvature} do not only  include   the flat $N$D Euclidean case, but also   the spherical, hyperbolic, Minkowskian and (anti-)de Sitter spaces (for the relativistic ones see~\cite{vulpiPAN, kiev}).

On the other hand, if the condition that all the integrals of the motion have to be quadratic in momenta is suppressed, the list of $N$D MS systems
is enlarged with the rational and hyperbolic Calogero--Sutherland--Moser models of type
$A_{N-1}$~\cite{Ca71,Su71,Mo75,Wo83}, the nonisotropic oscillator with rational frequencies~\cite{Jauch},
the nonperiodic Toda lattice~\cite{To67,ADS06}, the 
Benenti systems~\cite{Benentia, Benentib} and, very recently, the KC potential on the 3D Euclidean space plus {\em three} centrifugal terms~\cite{Verrier}.

From this viewpoint it is quite natural to wonder whether the number of $N$D QMS systems is much larger than the number of MS ones (obviously, any MS system is by construction a QMS one). The answer to this question has been recently answered affirmatively in~\cite{PLB} as follows: 
geodesic flows on {\em any} $N$D spherically symmetric (curved) space define always a QMS Hamiltonian. Moreover, all these systems present the {\em same} ``universal" and explicit set of $(2N-3)$ integrals in involution with the Hamiltonian that are quadratic in the momenta. 
The proof of this result   is based on the common $\rsl(2,\RR)$ Poisson coalgebra symmetry~\cite{BR98,Deform,CRMAngel}
 of all these systems, a fact that was also shown for the
quadratically MS systems on Riemannian spaces listed above~\cite{BH07, sigmaorlando, PhysD,PLB,BR98}.  

The aim of this paper is to use the same $\rsl(2,\RR)$-coalgebra symmetry in order to show that such infinite family of QMS free systems in arbitrary dimension can be further enlarged by including the Hamiltonians
\be
{\cal H}=\frac{1}{2f(|\bq|)^2}\left\{
\bp^2+ \frac{\la}{\bq^2}+\sum_{i=1}^N \frac{b_i}{q_i^2}\right\}+{\cal U}(|\bq|),
\qquad   (|\bq|=\sqrt{\bq^2}) ,
\label{hamm}
\ee
 where   $f$ and ${\cal U}$ are   smooth functions and $\mu,b_1,\dots,b_N$ are real constants. The proof that this Hamiltonian is indeed QMS is presented in section 2. We stress that, despite that the spherical symmetry is broken by the ``centrifugal terms" ${b_i}/{q_i^2}$, the overall $\rsl(2,\RR)$-coalgebra symmetry holds and, as a consequence, the {\em same} integrability properties of the geodesic flows are preserved when the potentials are considered. Next sections will be devoted to the study of certain particular cases of \eqref{hamm}, both from the physical and from the mathematical point of view.
 
Firstly, note that the term $\frac 12 {\bp^2}/{f(|\bq|)^2}$ is just the kinetic energy that defines the geodesic flow on a {\em spherically symmetric space}, which is generically of  {\em nonconstant curvature} 
($f$ playing the role of a  ``conformal factor" of the flat metric). These are just the aforementioned geodesic flows  on $N$D spherically symmetric spaces~\cite{PLB}. In particular, suitable choices for the function $f$ allow us to recover the three classical Riemannian spaces of constant curvature~\cite{Doub}, $N$D generalizations~\cite{PhysD, PLB} of the Darboux surfaces~\cite{Ko72, KKMW03} and the Iwai--Katayama spaces~\cite{IK95} (which are themselves a generalization of the Taub--NUT metric~\cite{IK93}). All these QMS geodesic flows are fully described in section 3 by using the $\rsl(2,\RR)$-coalgebra framework, where as a new result the complete integrability and separability of the free Hamiltonian (for any $f$)  is explicitly shown by using spherical coordinates.

Secondly,  section 4 is devoted to the study of the central potential $\cal U$ and to the analysis of  its separability properties which hinges on its Lie--Poisson coalgebra symmetry.  In section 5, the $N$D versions of the potential  $\cal U$ defining the ``intrinsic" KC and   oscillator  potentials on generic spherically symmetric spaces is presented by extending to arbitrary dimension (see~\cite{PhysD}) the definitions of such potentials on 3D Riemannian spaces previously introduced in~\cite{EP06d,EP08, LT87, LT95, CQG}. Moreover, by taking into account that  the  ${\la}/{(2f(|\bq|)^2  \bq^2)}$ contribution can be interpreted as a Dirac monopole-type term~\cite{IK95,  Br89, SW94, SW94b, CK01, KMOZ07, KN07, MNY07, NR07, GW07, NY08, Na08}  and  by adding  $N$  centrifugal  terms to the KC/oscillator potential ${\cal U}$, new  generalized KC  and oscillator systems on arbitrary spherically symmetric spaces are introduced.    
These general results are explicitly illustrated in section 6 by studying  in detail the resulting potentials for some specific spaces. In particular,    known superintegrability results for the MIC--Kepler~\cite{IK95,KN07, NY08,mica,micb,micc}   and  the  Taub-NUT systems~\cite{IK95, GW07,Ma82,AH85,GM86,FH87,GR88, BCJ, BCJM, JL} are recovered within our framework,  and we also derive further results for  new Hamiltonians such as the proper KC and oscillator potentials  on the $N$D Darboux spaces introduced in~\cite{PLB}.  This way new QMS systems on $N$D curved spaces are presented and analysed under a unified integrability setting, showing that the integrability properties of several  well-known systems rely on a common $\rsl(2,\RR)$-coalgebra symmetry. 
 

\section{The Hamiltonian ${\cal H}$ is QMS}

The following result holds:

 \noindent
{\bf Theorem 1.}
{\it Let $\{\>q,\>p \}=\{(q_1,\dots,q_N),(p_1,\dots,p_N)\}$ be $N$ pairs of
canonical variables. The $N$D Hamiltonian
\begin{equation}
{\cal H}=\frac{1}{2f(|\bq|)^2}\left\{
\bp^2+ \frac{\la}{\bq^2}+\sum_{i=1}^N \frac{b_i}{q_i^2}\right\}+{\cal U}(|\bq|),
\qquad   (|\bq|=\sqrt{\bq^2}),
\nonumber
\end{equation}
 where   $f$ and ${\cal U}$ are   smooth functions and $\mu,b_1,\dots,b_N$ are real constants, is 
quasi-maximally superintegrable. The   $(2N-3)$ functionally
independent and ``universal" integrals of the motion for ${\cal H}$ are given by
\bea
&& C^{(m)}= \sum_{1\leq i<j}^m \left\{ ({q_i}{p_j} -
{q_j}{p_i})^2 + \left(
b_i\frac{q_j^2}{q_i^2}+b_j\frac{q_i^2}{q_j^2}\right)\right\}
+\sum_{i=1}^m b_i ,\nonumber\\
&& C_{(m)}= \sum_{N-m+1\leq i<j}^N \left\{ ({q_i}{p_j} -
{q_j}{p_i})^2 + \left(
b_i\frac{q_j^2}{q_i^2}+b_j\frac{q_i^2}{q_j^2}\right)\right\}
+\sum_{i=N-m+1}^N b_i ,
\label{cinfm}
\eea
where $m=2,\dots, N$ and $C^{(N)}=C_{(N)}$. Moreover, the sets of $N$ functions
$\{{\cal H},C^{(m)}\}$ and $\{{\cal H},C_{(m)}\}$ $(m=2,\dots, N)$ are in
involution. 
}
\medskip

 \noindent
{\bf Proof.} Let us define the functions
\be
J_-=\bq^2 ,\qquad J_3=\bq\cdot\bp ,\qquad J_+=\bp^2+\sum_{i =1}^N \frac{b_i }
{q_i^{2}} ,\label{Jep}
\ee
where $$
\bq^2=\sum_{i=1}^N q_i^2,\qquad \bp^2=\sum_{i=1}^N p_i^2,\qquad \bq\cdot\bp=\sum_{i=1}^N q_i p_i,\qquad b_i\in\RR.
$$
It is immediate to check that these three functions close the $\rsl(2,\mathbb R)$ Poisson  coalgebra
\be
\{J_3,J_+\}=2J_+ ,\qquad \{J_3,J_-\}=-2J_- ,\qquad
\{J_-,J_+\}=4J_3.
\ee
Moreover, ${\cal H}$ can be written as the following function of the $\rsl(2,\mathbb R)$-coalgebra generators:
\be
{\cal H}=\frac{J_++\la J_-^{-1}}{2f(\sqrt{J_-})^2}+{\cal U}(\sqrt{J_-}) .
\ee
As a consequence of this coalgebra symmetry, ${\cal H}$ will Poisson-commute with all the integrals $C^{(m)}$ and $C_{(m)}$ that respectively come from the left
and right $m$th-coproducts of the Casimir function for $\rsl(2,\mathbb R)$. 
This coalgebra symmetry also ensures that the sets of
$N$ functions
$\{\cH,C^{(m)}\}$ and 
$\{\cH,C_{(m)}\}$ ($1<m\leq N$) are in involution (see~\cite{sigmaorlando,BR98,CRMAngel} for explicit proofs of all these statements). 

In this respect, we recall that {\em any} Hamiltonian
\begin{equation}
\cH=\cH(J_-,J_+,J_3)  , \label{hgen}
\end{equation}
defined as a smooth function on the ${\rsl}(2,\mathbb R)$ generators  is  always  QMS, and has the same set of  ``universal" integrals~\eqref{cinfm} that are obtained through the $m$D symplectic realization of the Casimir function of ${\rsl}(2,\mathbb R)$. This underlying coalgebraic structure can be interpreted as a generalization of the
 spherical symmetry,  which is recovered when all $b_i=0$.

Finally, note that the centrifugal terms ${b_i}/{ q_i^2}$ 
come from the fact that ${\rsl}(2,\mathbb R)$ allows us to add such a contribution in the corresponding symplectic realization of the $J_+$ generator. Therefore, from an algebraic viewpoint these centrifugal contributions are directly related to the kinetic energy term.


\section{Superintegrability of   geodesic flows}

The metric of   any $N$D spherically symmetric space $\cM$  can be written as
\begin{equation}\label{metric}
\dd s^2=f(|\bq|)^2\,\dd\bq^2=f(r)^2(\dd r^2+r^2\dd\Om^2) ,
\end{equation}
which is expressed in terms of the following coordinate systems: 
\begin{itemize}

\item Generic coordinates $\bq=(q_1,\dots,q_N)$
(with $\dd\bq^2 =\sum_{i=1}^N\dd q_i^2$), which in our approach we shall identify with those appearing in the symplectic representation (\ref{Jep}).

\item  Spherical coordinates with a radial-type variable $r=|\bq|\in \RR^+$ (which does not usually  coincide with the geodesic distance) and   $N-1$ ordinary angular variables $\theta_j\in[0,2\pi)$ $(j=1,\dots,N-1)$ with     $\dd\Om^2$  being the standard metric on the unit $(N-1)$D sphere
\begin{equation}
\dd\Om^2=\sum_{j=1}^{N-1}\dd\te_j^2\prod_{k=1}^{j-1}\sin^2\te_k .
\label{om}
\end{equation}
\end{itemize}

In both coordinate systems    $f(|\bq|)=f(r)$ is any  smooth function, usually interpreted as a conformal factor of the Euclidean metric $\dd s^2=\dd\bq^2$. We stress that only in the  Euclidean case $\bq$ can be interpreted as Cartesian coordinates.
The relations between both sets of coordinates
read
\begin{equation}
q_j=r \cos\te_{j}     \prod_{k=1}^{j-1}\sin\te_k ,\quad 1\leq j<N , \qquad 
q_N =r \prod_{k=1}^{N-1}\sin\te_k ,
\label{otrasmetrics}
\end{equation}
where hereafter any product $\prod_{l}^m$ such that $l>m$ is assumed to be equal to 1. The scalar curvature of (\ref{metric}) turns out to be
\begin{equation}\label{curv}
R=-(N-1)\,\frac{2f''(r)+2(N-1)r^{-1}f'(r)+(N-2)f'(r)^2}{f(r)^2} ,
\end{equation}
 where    $f'(r)=\dd f/\dd r $   and $f''(r)=\dd^2 f/\dd r^2$. Therefore, we are indeed dealing with $N$D spaces with nonconstant curvature.


\subsection{Free Hamiltonians}
 
The metric (\ref{metric}) provides the free Lagrangian,   which characterizes  the geodesic motion of a particle on $\cM$  in terms of the velocities $\dot {\>q}$ or $(\dot r,\dot \theta_j)$:
\be
{\cal T}=\frac 12 f(|\bq|)^2  \dot {\>q}^2=\frac 12 f(r)^2\left(\dot r^2+r^2\sum_{j=1}^{N-1}\dot\te_j^2\prod_{k=1}^{j-1}\sin^2\te_k\right ) .
\label{lag}
\ee
Then the canonical momenta  $\>p, p_r,p_{\te_j}$,  conjugate to  $\>q, r, \te_j$, can be obtained through a Legendre transformation ($p_i=\partial {\cal T}/\partial\dot q_i$)
\be
{\>p}=f(|\bq|)^2  \dot {\>q} , \qquad
p_r=f(r)^2\dot r ,\qquad p_{\te_j}=f(r)^2 r^2 \dot\te_j \prod_{k=1}^{j-1}\sin^2\te_k .
\label{op}
\ee
From (\ref{otrasmetrics}) and  (\ref{op}) the relations between both sets of momenta $\>p$ and $p_r,p_{\te_j}$ are found to be
\bea
&& p_j=\prod_{k=1}^{j-1}\sin\te_k\cos\te_j\,p_r+\frac{\cos\te_j}{r}\sum_{l=1}^{j-1} \frac{\prod_{k=l+1}^{j-1}\sin\te_k }{\prod_{m=1}^{l-1}\sin\te_m} \cos\te_l\,p_{\te_l}-\frac{\sin\te_j}{r\prod_{k=1}^{j-1}\sin\te_k}\,p_{\te_j} , \nonumber\\
&&p_N=\prod_{k=1}^{N-1}\sin\te_k \,p_r+\frac 1 r\sum_{l=1}^{N-1}\frac{\prod_{k=l+1}^{N-1}\sin\te_k}{\prod_{m=1}^{l-1}\sin\te_m}\cos\te_l \, p_{\te_l},
\label{oq}
\eea
where $1\le j<N$ and from now on any sum   $\sum_{l}^m$ such that $l>m$ is assumed to be zero. Both sets of phase spaces $(\>q,\>p)$ and $(r,\te_j;p_r,p_{\te_j})$ are canonical coordinates and momenta with respect to the usual Poisson bracket
\bea
&&\{F,G\}=\sum_{i=1}^N\left(\frac{\partial F}{\partial q_i}
\frac{\partial G}{\partial p_i}
-\frac{\partial G}{\partial q_i} 
\frac{\partial F}{\partial p_i}\right)\nonumber\\
&&\qquad \quad\,=
\frac{\partial F}{\partial r}
\frac{\partial G}{\partial p_r}
-\frac{\partial G}{\partial r} 
\frac{\partial F}{\partial p_r}+
\sum_{j=1}^{N-1}\left(\frac{\partial F}{\partial \te_j}
\frac{\partial G}{\partial p_{\te_j}}
-\frac{\partial G}{\partial \te_j} 
\frac{\partial F}{\partial p_{\te_j}}\right) .
\label{bg}
\eea

This implies that the geodesic motion is described through the kinetic energy Hamiltonian
  \begin{equation}
{\cal T} = \frac{\bp^2}{2f(|\bq|)^2}=\frac{p_r^2+r^{-2}\bL^2}{2f(r)^2} ,
\label{or}
\end{equation}
where $\bL^2$ is the total   ``angular momentum", namely,
\begin{equation}
\bL^2=\sum_{j=1}^{N-1}p_{\te_j}^2\prod_{k=1}^{j-1}\frac{1}{\sin^{2}\te_k}.
\label{angular}
\end{equation}
Therefore   the kinetic term on  $\cM$ arises as an  $\rsl(2,\RR)$-coalgebra Hamiltonian (\ref{hgen}):
\be
{\cal T} =\frac{J_+}{2f(\sqrt{J_-})^2} ,
\label{os}
\ee
provided that {\em all} $b_i=0$. Therefore, this geodesic flow is always QMS for any function $f$ and its  constants of motion are just (\ref{cinfm}). 

We remark that in the case of free motion  we recover the well-known $\frak{so}(N)$  Lie--Poisson symmetry of spherically symmetric spaces. Explicitly, the functions $J_{ij}={q_i}{p_j} - {q_j}{p_i}$ with $i<j$ and $i,j=1,\dots,N$ span an $\frak
{so}(N)$ Lie--Poisson algebra  
\begin{equation}
\{ J_{ij},J_{ik} \}= J_{jk} ,\qquad  \{ J_{ij},J_{jk} \}= -J_{ik} ,\qquad 
\{ J_{ik},J_{jk} \}= J_{ij} , \qquad i<j<k .
\label{ot}
\end{equation}
Hence   the integrals (\ref{cinfm})  can be expressed as sums of such ``angular momentum"
components 
\begin{equation}
C^{(m)}= \sum_{1\leq i<j}^m J_{ij}^2 ,
\qquad C_{(m)}=\!\!\! \sum_{N-m+1\leq i<j}^N J_{ij}^2 .
\label{ov}
\end{equation}
Notice that $ C^{(N)}=C_{(N)}$   is the quadratic Casimir of $\frak
{so}(N)$ and, in fact, each $C^{(m)}$ (or $C_{(m)}$) is the quadratic Casimir of certain rotation subalgebra 
$\frak {so}(m)\subset \frak {so}(N)$.

By taking into account  (\ref{otrasmetrics}) and (\ref{oq}) we obtain  the generators $J_{ij}$ written  in terms of the spherical phase space variables; namely,
\bea
&&J_{ij}=\sin\te_i\cos\te_j\prod_{k=i}^{j-1}\sin\te_k\, p_{\te_i}-\frac{\cos\te_i\sin\te_j}{\prod_{k=i}^{j-1}\sin\te_k}\,  p_{\te_j}\nonumber\\
&&\qquad\quad  +\cos\te_i\cos\te_j \sum_{l=i}^{j-1}\frac{\prod_{k=l+1}^{j-1}\sin\te_k  }{\prod_{m=i}^{l-1}\sin\te_m}\cos\te_l \, p_{\te_l} , \qquad 1\le i<j<N ,\nonumber\\
&&J_{iN}=\sin\te_i \prod_{k=i}^{N-1}\sin\te_k\, p_{\te_i}+\cos\te_i\sum_{l=i}^{N-1}\frac{\prod_{k=l+1}^{N-1}\sin\te_k  }{\prod_{m=i}^{l-1}\sin\te_m}\cos\te_l\, p_{\te_l}  .
\label{ow}
\eea
By using these expression one can readily write down the constants of motion (\ref{ov}) in this set of canonical variables. The resulting   formulas for $C_{(m)}$ adopt the very compact form
\begin{equation}
C_{(m)}=\sum_{j=N-m+1}^{N-1}p_{\te_j}^2\prod_{k=N-m+1}^{j-1}\frac 1{\sin^{2}\te_k  } ,
\label{ox}
\end{equation}
so that $C_{(N)} =\bL^2$ is the    total   ``angular momentum" (\ref{angular}). 
Furthermore, the
complete integrability determined by the set of $N$ functions $\{{\cal T},C_{(m)}\}$
$(m=2,\dots, N)$ leads   to a separable
  system as a set of $N$ equations each of them depending only on a pair of
canonical quantities. In particular  we obtain a  {\em common} set of $N-1$  angular equations for the   free Hamiltonian
(\ref{or})  
\bea
&& C_{(2)}(\te_{N-1},p_{\te_{N-1}})=p^2_{\te_{N-1}} , \nonumber\\[2pt]
&& C_{(l)}(\te_{N-l+1},p_{\te_{N-l+1}})=p^2_{\te_{N-l+1}} +
\frac{C_{(l-1)} }{\sin^{2}\te_{N-l+1}} , \qquad l=3,\dots,N-1, \nonumber \\[2pt]
&& C_{(N)}(\te_{1},p_{\te_{1}})=p^2_{\te_{1}} +
\frac{C_{(N-1)} }{\sin^{2}\te_1}\equiv  \bL^2  ,
\label{mc}
\eea
together with a single  radial equation corresponding to the Hamiltonian itself
${\cal T}(r,p_r)$.


\subsection{Some examples of QMS geodesic flows}

The  $\rsl(2,\RR)$-coalgebra setting  describes all spherically symmetric spaces and, consequently, many geodesic motion Hamiltonians that have been studied in the literature by following different procedures can be embedded into this integrability framework.  In what follows  we shall present   three families of examples whose explicit metric and free Hamiltonian are displayed in table~\ref{table1}.

\begin{table}[h]
{\footnotesize
 \noindent
\caption{{Geodesic flow Hamiltonians with  $\rsl(2,\RR)$-coalgebra symmetry for some particular $N$D spherically symmetric spaces in the generic coordinates $(\>q,\>p)$.}}
\label{table1}
\medskip
\noindent\hfill
$$
\begin{array}{lll}
\hline
\\[-6pt]
\mbox{Space}&  \mbox{Metric} \  \dd s^2 &   \mbox{Hamiltonian}\ {\cal T}={\cal T}(J_-,J_+)\\[4pt]
 \hline
  \\[-6pt]
\mbox{Euclidean}&   \dd \>q^2  &\frac12 J_+  \\[6pt]
\mbox{Spherical}& \displaystyle { \frac { 4\, \dd   \>q^2}
{(1+\>q^2)^2 } } & {\frac12 (1+J_-)^2 J_+    }\\[8pt]
\mbox{Hyperbolic}& \displaystyle { \frac { 4\, \dd   \>q^2}
{(1-\>q^2)^2 } } & {\frac12 (1-J_-)^2 J_+    }\\[12pt]
 \hline
  \\[-6pt]
\mbox{Darboux I}&  \displaystyle {  \frac{\ln|\bq|\,\dd\bq^2}{\bq^2}}  & \displaystyle{\frac{J_- J_+}{2\ln\sqrt{J_-}}  }\\[10pt]
\mbox{Darboux II}&  \displaystyle {   \frac{1+ \ln^{2}|\bq| }{\bq^2\ln^{2}|\bq|}\, \dd\bq^2 }  & \displaystyle{\frac{J_- \ln^2\sqrt{J_-} \,J_+}{2(1+\ln^2\sqrt{J_-})}  }\\[10pt]
\mbox{Darboux IIIa}&  \displaystyle {      \frac{1+|\bq|}{\bq^4}\,\dd\bq^2 }  & \displaystyle{\frac{J_-^2 J_+}{2(1+\sqrt{J_-})} 
  }\\[8pt]
\mbox{Darboux IIIb}&  \displaystyle {  (\kk+\bq^2)\,\dd\bq^2 }  & \displaystyle{ \frac{J_+}{2(\kk+J_-)} 
  }\\[10pt]
\mbox{Darboux IV}$\qquad$ &  \displaystyle {\frac{a+\cos(\ln|\bq|)}{\bq^2\sin^2(\ln|\bq|)}\,\dd\bq^2 \qquad}  & \displaystyle{\frac{J_-\sin^2(\ln\sqrt{J_-})J_+}{2(a+\cos(\ln\sqrt{J_-}))} 
  }\\[12pt]
 \hline
  \\[-6pt]
\mbox{Taub-NUT}&   \displaystyle {  \frac{(4m+|\bq|)\,\dd\bq^2}{|\bq|}}  & \displaystyle{\frac{\sqrt{J_-} J_+}{2(4m+\sqrt{J_-})}    }\\[10pt]
\mbox{$\nu$-fold $a\ne 0$}&  \displaystyle {  \frac{ (a+b |\bq|^{\frac 1\nu})\,\dd\bq^2}{   |\>q|^{2-\frac 1{\nu}}   }}  & \displaystyle{
\frac{ J_-^{1-\frac{1}{2\nu}} J_+}{2(a+b J_-^{\frac 1{2\nu}})}    }\\[10pt]
\mbox{$\nu$-fold $a= 0$}&   \displaystyle {   |\bq|^{\frac 2\nu - 2}\,\dd\bq^2  }  & \displaystyle{
\frac 12     J_-^{1-\frac{1}{\nu}} J_+   }\\[14pt]
 \hline
 \end{array}
$$
\hfill}
\end{table}

\begin{itemize}
\item  {\em The three simply connected Riemannian spaces of constant curvature}.
If we take $f(|\bq|)=2/(1+\k \bq^2)$ we recover the spherical $(\k>0)$, Euclidean $(\k=0)$ and hyperbolic $(\k<0)$ spaces  of constant sectional curvature $\k$ (the scalar one is $N(N-1)\k$). The generic coordinates $\bq$ are identified with Poincaré coordinates~\cite{Doub} coming from a stereographic projection in $\RR^{N+1}$. Note that    $r=|\bq|$ is not a geodesic distance; in fact, the proper geodesic   coordinate $\hat r$  is related to $r$ by 
\be
r=\frac{1}{\sqrt{\k}} \tan\left(\sqrt{\k}\,\frac {\hat r}2\right) ,\qquad
\dd s^2=\dd {\hat r}^2+\frac{\sin^2(\sqrt{\k}\hat r)}{\k}\,\dd \Om^2.
\label{geod}
\ee
As it is well-known 
the geodesic motion on these spaces is MS and their additional quadratic integral of motion in Poincaré coordinates  can be found  in~\cite{BH07}.  In table~\ref{table1} we write the resulting expressions for $\k=\{± 1,0\}$.

\item  {\em ND  Darboux spaces}.   The four  2D  {Darboux spaces}
are  surfaces of nonconstant curvature whose geodesic motion is quadratically MS~\cite{Ko72, KKMW03}. From the point of view of superintegrability, these surfaces can be seen as the closest analogues of the Riemannian surfaces of constant curvature. An $N$D spherically symmetric generalization for them was recently proposed in~\cite{PLB} by requiring them to be endowed with  an $\rsl(2,\RR)$-coalgebra symmetry. We remark that the Darboux surface of type III admits two (equivalent) generalizations of this kind:   the    type IIIa given in~\cite{PLB} and the type IIIb constructed in~\cite{PhysD}.  Only for this Darboux III space the MS property has been proven in arbitrary dimension~\cite{PhysD} (and the additional  integral is also quadratic), while for the three remaining types the MS problem is still open. Recall also that the   parameters $a$ and $k$ appearing in the types IIIb and IV in table~\ref{table1} are real constants.

 \item  {\em Iwai--Katayama spaces}.  These are the $N$D counterpart of the 3D spaces  
 underlying the so called  ``multifold Kepler" systems introduced by these authors in~\cite{IK95}. The generic $\nu$-fold Kepler space depends on two real constants, $a$ and $b$, as well on a {\em rational} parameter $\nu$ as shown in table~\ref{table1}.   For the sake of a further more transparent discussion we have distinguished   the cases $a\ne 0$ and $a=0$, and in the latter case $b$ can always be taken equal to 1.  The  Iwai--Katayama systems are of physical interest as they are  generalizations of the MIC--Kepler and the Taub-NUT  ones. As far as the underlying  space is concerned the Taub-NUT 
 metric   is recovered from the $\nu$-fold Kepler one provided that $\nu=1$,  $a=4m$ and $b=1$  (recall that the MIC--Kepler space is just the Euclidean one).
  We stress that in the 3D case all these Hamiltonians have been shown to be MS, but  the additional integral of motion is not, in general, quadratic in the momenta (see~\cite{CMP} for a detailed discussion on the subject).

\end{itemize}

It is also worth stressing that in 3D the classical Riemannian spaces, the Darboux space of type IIIb and the Iwai--Katayama spaces  are just particular cases of the so-called ``Bertrand spaces". The $N$D version of the whole family of 3D Bertrand spaces has been recently studied in~\cite{CQG, CMP} by making use of  the same $\rsl(2,\RR)$-coalgebra symmetry, and their explicit expressions for the abovementioned particular cases are already included in table~\ref{table1}. For the sake of simplicity, we have  omitted the generic form of the Bertrand metrics in table~\ref{table1}, but their $N$D generalization only requires to consider the appropriate conformal factor coming from~\cite{CQG, CMP} and to replace the metric $\dd\Om^2$ (and its associated 2D angular momentum $\>L^2$) on the 2D sphere by the metric (\ref{om}) on the $(N-1)$D sphere (and associated $(N-1)$D ``angular momentum" (\ref{angular})).


\sect{Superintegrable potentials on curved spaces}

 At this point, the  Hamiltonian  (\ref{hamm}) can be thought of as the kinetic energy (\ref{or})  plus  a $\la$-term and  a function ${\cal U}(|\bq|)$, and by considering a symplectic realization with non-vanishing $b_i$'s. This system   can also be expressed in terms of the   spherical phase
space variables (\ref{otrasmetrics}) and (\ref{oq}) by taking into account that the generators
$\{J_±,J_3\}$  turn out to be
\begin{equation}
\begin{aligned}
J_- &= r^2  , \qquad J_3= r p_r  ,\\
J_+ &= p_r^2+r^{-2}\bL^2+\sum_{j=1}^{N-1}\frac{b_j^2}{r^2 \cos^2\te_{j}    
\prod_{k=1}^{j-1}\sin^2\te_k}+\frac{b_N}{r^2\prod_{k=1}^{N-1}\sin^2\te_k} .
\end{aligned}
\label{otrareal}
\end{equation}
In this way we find that  the Hamiltonian (\ref{hamm}) can be rewritten as
\bea
 &&{\cal H}=\frac{J_+ +\la J_-^{-1}}{2f(\sqrt{J_-})^2}+{\cal U}(\sqrt{J_-}) =    \frac{p_r^2+r^{-2}\bL^2}{2f(r)^2}+\frac{\la}{2 r^2f(r)^2 }+{\cal U}(r)                \nonumber\\[2pt]
&&\qquad  +
\frac 1{2f(r)^2} \sum_{j=1}^{N-1}\frac{b_j^2}{r^2 \cos^2\te_{j}    
\prod_{k=1}^{j-1}\sin^2\te_k}+\frac 1{2f(r)^2} \frac{b_N}{r^2\prod_{k=1}^{N-1}\sin^2\te_k}.
\label{hammm}
\eea
Again, we stress that the symplectic realization of the generator $J_+$ is the essential tool    to
 incorporate in a natural way up to
$N$   centrifugal potentials associated to the  $b_i$'s.  

Therefore this Hamiltonian   is 
  QMS as it is endowed
with the $(2N-3)$ integrals of motion given by (\ref{cinfm}). However, the existence of an
additional integral providing  the MS property for $\cal H$ is, in general, not
guaranteed. Nevertheless, as it already happened with the free motion, ${\cal H}$ is separable in the latter form (\ref{hammm}). In this case the integrals of motion $C_{(m)}$ (\ref{cinfm}), with arbitrary centrifugal terms,  are given by
\begin{equation}
C_{(m)}=\sum_{j=N-m+1}^{N-1}\left( p_{\te_j}^2+\frac{b_{j}} {\cos^2\te_j} \right) \prod_{k=N-m+1}^{j-1} \frac 1{ \sin^{2}\te_k  } +\frac{b_N}{ \prod_{l=N-m+1}^{N-1}  \sin^{2}\te_l   }.
\label{oxx}
\end{equation}
Consequently, we obtain again  a  {\em common} set of $N-1$ {\em angular} equations
 \bea
&& C_{(2)}(\te_{N-1},p_{\te_{N-1}})=p^2_{\te_{N-1}}+ \frac{b_{N-1}}{\cos^2\te_{N-1}}+\frac{b_{N}}{\sin^2\te_{N-1}} , \nonumber\\[2pt]
&& C_{(l)}(\te_{N-l+1},p_{\te_{N-l+1}})=p^2_{\te_{N-l+1}} +
\frac{C_{(l-1)} }{\sin^{2}\te_{N-l+1}} +   \frac{b_{N-l+1}}{\cos^2\te_{N-l+1}}  ,\quad l=3,\dots,N-1, \nonumber \\[2pt]
&& C_{(N)}(\te_{1},p_{\te_{1}})=p^2_{\te_{1}} +
\frac{C_{(N-1)} }{\sin^{2}\te_1} +  \frac{b_{1}}{\cos^2\te_{1}}  \nonumber\\[2pt]
&&\qquad\qquad \quad\ \, =   \bL^2 +\sum_{j=1}^{N-1}\frac{b_j^2}{\cos^2\te_{j}    
\prod_{k=1}^{j-1}\sin^2\te_k}+\frac{b_N}{\prod_{l=1}^{N-1}\sin^2\te_l} ,
\label{mcc}
\eea
(to be compared with (\ref{mc}))
plus {\em one  radial} equation which depends on the specific Hamiltonian under consideration
\be
{\cal H}(r,p_r)=\frac{p_r^2+ r^{-2} C_{(N)} + r^{-2} \la}{2f(r)^2} +{\cal U}(r) .
\label{me}
\ee
Thus   when any $b_i\ne 0$ the spherical symmetry is broken and the ``total angular momentum" $\bL^2$ (\ref{angular}) is no longer a constant of motion, its role being now played by $C_{(N)}$ (\ref{mcc}) (only when all $b_i=0$, $C_{(N)}$  reduces to $\bL^2$). In the case with $b_i\ne 0$,  the generator $J_+$ (\ref{otrareal}) simply reads
\be
J_+=p_r^2+r^{-2}C_{(N)} ,
\label{mmee}
\ee
and the Poisson brackets among the $\rsl(2,\RR)$ generators (\ref{otrareal}) and $\bL^2$ turn out to be
\bea
&&\{J_-,\bL^2\}=\{J_3,\bL^2\}=0 ,\nonumber
\\[2pt]
&&\{J_+,\bL^2\}= \frac{4}{r^2}\sum_{j=1}^{N-1}\frac{\sin^2\te_j\tan\te_j \,p_{\te_j}}{\prod_{k=1}^{j-1}\sin^2\te_k}\nonumber\\[2pt]
&&\qquad × \left( b_j\tan^4\te_j-\sum_{l=j+1}^{N-1} \frac{b_l}{\cos^2\te_l \prod_{k=j+1}^{l-1}\sin^2\te_k}-\frac{b_N}{\prod_{s=j+1}^{N-1}\sin^2\te_s}\right) .
\label{mf}\eea

Summing up,   once a single nonzero parameter $b_i$ is allowed, the spherical symmetry is broken,
and the chain of rotation subalgebras  $\frak {so}(m)\subset \frak {so}(N)$   do not provide symmetries for the Hamiltonian, as it was the case of the Casimirs (\ref{ov})   for the free motion. Therefore, the $\mathfrak{sl}(2,\RR)$ Lie--Poisson coalgebra symmetry can be understood as the appropriate generalization of the usual rotational symmetry which allows to deal with the general case above.

We also remark, as pointed out in the introduction,  that the $\la$-potential in (\ref{me}) (or (\ref{hammm}))  can be interpreted as a  Dirac  monopole-type term~\cite{IK95, KN07, MNY07, NY08}. 
Hence  since   $C_{(N)}$ is an  integral of the motion, the appearance of this monopole-type potential can be seen, from a dynamical viewpoint, as  coming from a radial centrifugal term ruled by  $C_{(N)}$. In particular, when there are no  centrifugal potentials $(b_i=0)$, the $\la$-term has already been interpreted, for some concrete systems, as a {\em proper} Dirac monopole ~\cite{IK95,  Br89, SW94, SW94b, CK01, KMOZ07, KN07, MNY07, NR07, GW07, NY08, Na08}
   that actually comes from the  ``total angular momentum"  through   $C_{(N)}\equiv \bL^2\to \bL^2+\la$.


\section{KC and oscillator potentials on curved spaces}

Among all the possible choices of the central function ${\cal U}(r)$, two of them should correspond to the 
appropriate $N$D  definition of the  intrinsic KC and oscillator potentials on $\cM$. As we shall see in the sequel, a solution to this problem can be found through a suitable generalization of the 3D construction for such potentials.

  Let us consider the  Laplace--Beltrami operator on the $3$D   spherically symmetric manifold
  $\cM^3$ with metric (\ref{metric}) in the coordinates
$\bq$:
\be
\De_{\cM^3}=\sum_{i,j=1}^3\frac1{\sqrt g}\dif{}{q_i}\sqrt gg^{ij}\dif{}{q_j} ,
\label{wa}
\ee
where 
$g^{ij}$ is the   inverse  of the   metric tensor   $g_{ij}$ and $g$  is the determinant of $g_{ij}$.
  The radial symmetric Green function $U(|\bq|)=U(r)$ on $\cM^3$ (up to multiplicative and additive
constants) is defined as the positive nonconstant solution to the equation
\be
\De_{\cM^3}U(r)=
\frac{ 1}{r^2 f(r)^3}\frac\dd{\dd
r}\left(r^2 f(r)\frac{\dd U(r)}{\dd
r}\right) =0
\quad\text{on}\quad \cM^3\minus\{\textbf0\}\, ,
\label{wb}
\ee
so that
\begin{equation}
U(r) =\int^r\frac{\dd r'}{r'^2f(r')} .
\label{wc}
\end{equation}
Now, the essential point in the $N$D case~\cite{PhysD}  is the fact that we can keep the very same definitions given in~\cite{EP06d,EP08, LT87, LT95, CQG}  for the KC and oscillator potentials on a 3D spherically symmetric space. In particular, the
\emph{intrinsic KC potential}  on  the $N$D space $\cM$ will be defined by
\begin{equation}
{\cal U}\Kep(r) :=\cte\,U (r) ,
\label{wd}
\end{equation}
while the
\emph{intrinsic   oscillator potential}  is defined to be proportional
to the inverse square of the KC potential
\begin{equation}
{\cal U}\Harm(r) :=\frac \cteb{U^2(r)} ,
\label{we}
\end{equation}
where  $\cte$ and $\cteb$ are     real constants.

Clearly, the intrinsic KC and oscillator potentials do preserve the $\rsl(2,\RR)$-coalgebra symmetry since they are defined through a function of $|\bq|=r=\sqrt{J_-}$. As a straightforward  consequence, such a  symmetry   allows us to
 give the    superposition of either the  intrinsic  KC (\ref{wd}) or  
oscillator (\ref{we}) potential with  $N$ centrifugal terms and also with a monopole-type term. In
terms of the   symplectic realization (\ref{Jep}) and by considering arbitrary parameters
$b_i$, the curved KC system read
\be
{\cal H}\sub{KC}:=\frac{J_+ +\la J_-^{-1}}{2f(\sqrt{J_-})^2}+{\cal U}\sub{KC}(\sqrt{J_-})
=\frac{\bp^2+\la \bq^{-2} +\sum_{i=1}^N b_i q_i^{-2}}{2 f(|\bq|)^2}+\cte\,{U}(|\bq|)  
\label{wf}
\ee
and the intrinsic oscillator on a curved space is defined as
\be
{\cal H}\sub{O}:=\frac{J_+ +\la J_-^{-1}}{2f(\sqrt{J_-})^2}+{\cal U}\Harm(\sqrt{J_-})
 =\frac{\bp^2+\la \bq^{-2}+\sum_{i=1}^N b_i q_i^{-2}}{2f(|\bq|)^2}+\frac \cteb{U(|\bq|)^2} .
\label{wg}
\ee
Both systems can be immediately written in terms of the spherical phase space variables by means of (\ref{otrareal}).


\section{Some physically relevant examples}

The  results of the previous section are explicitly  illustrated by writing in table~\ref{table2} the QMS  Hamiltonians (\ref{wf}) and (\ref{wg}) associated to all the particular spaces  given in table~\ref{table1}. For each $N$D curved  space  we display  the   KC,  oscillator,  monopole-type and centrifugal potentials (with parameters $\cte$, $\cteb$, $\la$ and $b_i$, respectively). A specific QMS Hamiltonian $\cal H$ thus comes from the superposition of some (or all) of the corresponding terms.  Notice that, in some cases, the  multiplicative constants 
that arise when computing the Green function (\ref{wc}) have been absorbed within  $\cte$ and $\cteb$. 

In what follows we explicitly analyse  these systems for the three families of spaces discussed in section 3.2, since they are indeed the most relevant ones from the physical point of view.

\begin{table}[ht]
{\footnotesize
 \noindent
\caption{{Intrinsic KC,   oscillator,  monopole-type  and centrifugal potentials  with  $\rsl(2,\RR)$-coalgebra symmetry in the generic coordinates $\bq$ corresponding to the free Hamiltonians on the particular $N$D spherically symmetric spaces  given in table~\ref{table1}.}}
\label{table2}
\medskip
\noindent\hfill
$$
\begin{array}{lllll}
\hline
\\[-6pt]
\mbox{Space}& \mbox{KC}& \mbox{Oscillator} &   \mbox{Monopole}  &   \mbox{Centrifugal terms}\\[4pt]
 \hline
  \\[-6pt]
\mbox{Euclidean}& \displaystyle {-\frac \cte  {|\bq|}}& \cteb\,\bq^2  & \displaystyle{\frac{\la}{2 \bq^2} }&\displaystyle \frac 12 {\sum_{i=1}^N\frac{b_i}{q_i^2}} \\[6pt]
\mbox{Spherical}&\displaystyle {\frac{\cte(\bq^2-1)}{|\bq|} }  & \displaystyle {  \frac { \cteb\,\bq^2}
{(\bq^2-1)^2 } }& \displaystyle{ \frac{\la  (1+\bq^2)^2  }{2\bq^2}  } & \displaystyle \frac 12(1+\bq^2)^2 {\sum_{i=1}^N\frac{b_i}{q_i^2}} \\[8pt]
\mbox{Hyperbolic}&\displaystyle {-\frac{\cte(\bq^2+1)}{|\bq|} }  & \displaystyle {  \frac { \cteb\,\bq^2}
{(\bq^2+1)^2 } } & \displaystyle{ \frac{\la  (1-\bq^2)^2  }{2\bq^2}  }& \displaystyle \frac 12(1-\bq^2)^2 {\sum_{i=1}^N\frac{b_i}{q_i^2}} \\[12pt]
 \hline
  \\[-6pt]
\mbox{Darboux I}& \displaystyle{ \cte\sqrt{\ln |\>q|  }} & \displaystyle {  \frac{\cteb}{ \ln |\>q| }}  &   \displaystyle{\frac{\la}{2\ln|\bq|}}&
  \displaystyle \frac {\bq^2}{2\ln|\bq|} {\sum_{i=1}^N\frac{b_i}{q_i^2}}\\[10pt]
\mbox{Darboux II}& \displaystyle{\cte\sqrt{1+ \ln^2 |\>q|  }}& \displaystyle {  \frac{\cteb}{ 1+\ln^2 |\>q| }}  &
\displaystyle {\frac{\la\ln^2|\bq|}{2(1+\ln^2|\bq|)} }& \displaystyle \frac {\bq^2\ln^2|\bq|}{2(1+\ln^2|\bq|)} {\sum_{i=1}^N\frac{b_i}{q_i^2}}\\[10pt]
 \mbox{Darboux IIIa }& \displaystyle{\cte\sqrt{1+   |\>q|  }}& \displaystyle {  \frac{\cteb}{ 1+  |\>q| }}  &
\displaystyle {\frac{\la \bq^2}{2(1+|\bq|)} }&   \displaystyle \frac {\bq^4}{2(1+ |\bq|)} {\sum_{i=1}^N\frac{b_i}{q_i^2}}\\[10pt]
 \mbox{Darboux IIIb}& \displaystyle{ \frac{\cte\sqrt{\kk+\bq^2 }}{|\bq|}} & 
\displaystyle{ \frac{\cteb\, \bq^2  }{\kk+\bq^2}} & 
\displaystyle {\frac{\la}{2\bq^2 (\kk+\bq^2) } }&    \displaystyle \frac {1}{2(\kk+ \bq^2)} {\sum_{i=1}^N\frac{b_i}{q_i^2}}\\[10pt]
 \mbox{Darboux IV}& \displaystyle{\cte   { \sqrt{a+\cos(\ln|\bq|)} } } & \displaystyle {      \frac{\cteb}{a+\cos(\ln|\bq|)}  }&\displaystyle {            \frac{\la \sin^2(\ln|\bq|)}{2  (a+\cos(\ln|\bq|))} }   & \displaystyle{ \frac{\bq^2\sin^2(\ln|\bq|)}{2(a+\cos(\ln|\bq|))} 
\sum_{i=1}^N\frac{b_i}{q_i^2} }\\[12pt]
 \hline
  \\[-6pt]
  \mbox{Taub-NUT}& \displaystyle{\cte\sqrt{4m |\bq|^{-1} +1} } & \displaystyle {  \frac{ \cteb  |\bq|}{4m  + |\bq|}}&\displaystyle {\frac{\la}{   2|\bq| (4m+|\bq| ) } }& \displaystyle{ 
   \frac{   |\>q| } { 2(4m+ |\bq|)  }  \sum_{i=1}^N\frac{b_i}{q_i^2}  } \\[10pt]
\mbox{$\nu$-fold $a\ne 0$}& \displaystyle{\cte\sqrt{a |\bq|^{-\frac 1\nu} +b} } & \displaystyle {  \frac{ \cteb}{a |\bq|^{-\frac 1\nu} +b}}  &  \displaystyle {    \frac{\la}  {2|\bq|^{\frac 1\nu} (a+b |\bq|^{\frac 1\nu}) }   }&   \displaystyle{ 
   \frac{   |\>q|^{2-\frac 1{\nu}}} { 2(a+b |\bq|^{\frac 1\nu}) }  \sum_{i=1}^N\frac{b_i}{q_i^2}  }\\[10pt]
\mbox{$\nu$-fold $a= 0$}& \displaystyle{-\frac {\cte}{  |\bq|^{\frac 1\nu}}  } & \displaystyle {   \cteb |\bq|^{\frac 2\nu}}  &  \displaystyle {    \frac{\la}  {2|\bq|^{\frac 2\nu} }   }&   \displaystyle{ 
   \frac 12 \,  |\>q|^{2-\frac 2{\nu}}     \sum_{i=1}^N\frac{b_i}{q_i^2}  }\\[14pt]
\hline
 \end{array}
$$
\hfill}
\end{table}


\subsection {The   classical Riemannian spaces of constant curvature}

For any simply connected Riemannian  space with constant sectional curvature $\k$,  the metric function $f$ in Poincaré coordinates~\cite{BH07} reads
$$
f(|\bq|)=  \frac{ 2 }{1+\k\,\bq^2},
$$
and the Green function \eqref{wc} is easily computed:
$$
U(|\bq|)=  \frac{ \k\bq^2-1}{|\bq|}.
$$
As a consequence, the KC and oscillator potentials are the ones shown in table~\ref{table2} for $\k=\{± 1,0\}$. Such expressions can be rewritten in a more usual form by introducing the geodesic radial coordinate $\hat r$ (\ref{geod}) which gives
$$
{\cal U} \Kep(\hat r)=-\cte \, \frac{\sqrt{\k}}{\tan(\sqrt{\k}\hat r)} , \qquad
{\cal U} \Harm(\hat r)=\cteb\, \frac{\tan^2(\sqrt{\k}\hat r)}{\k} .
$$
We recall that when all the
$N$ parameters $b_i$ are diferent from zero, the generalized KC Hamiltonian  is QMS but not quadratically MS. Moreover, only when at
least one of the $b_i$'s vanishes an additional quadratic integral of motion arises~\cite{BH07} (which is a component of the Laplace--Runge--Lenz
$N$-vector). Nevertheless,  it has been recently proven in~\cite{Verrier} that 
when the KC potential is constructed on the 3D Euclidean space it is possible to consider the {\em three} centrifugal potentials obtaining a MS Hamiltonian with an additional {\em quartic} integral.   In contrast, the so-called Smorodinsky--Winternitz system ({\em i.e.}, the oscillator plus $N$ centrifugal terms)  is always MS for any value of the $b_i$'s.

We stress that the  MIC--Kepler Hamiltonian~\cite{IK95,KN07, NY08,mica,micb,micc} 
is also recovered when the KC and the monopole potential are considered in the flat Euclidean space:
\be
{\rm Euclidean:}\quad {\cal H}_{\mathrm {MIC-Kepler}}=\frac 12 J_+ -\frac{\cte}{\sqrt{J_-}} +\frac{\la}{2J_-}=
\frac12 \>p^2 -\frac \cte  {|\bq|}+  \frac{\la}{2 \bq^2}  .
\label{wh}
\ee
Therefore, as a byproduct of our procedure, the curved MIC--Kepler   analogue on the $N$D spherical and hyperbolic spaces can be constructed from  table~\ref{table2}. Namely,
\bea
\!\!\!\! \!\!\!\!  {\rm Spherical:} &&\!\!\!\! {\cal H}_{\mathrm {MIC-Kepler}}=\frac12 (1+J_-)^2 J_+  +\frac{\cte(J_- -1)}{\sqrt{J_-}} +\frac{\la(1+J_-)^2}{2J_-}\nonumber\\[2pt]
&&\qquad\qquad \ \ =
\frac 12{(1+\>q^2)^2 }  \>p^2  + \frac{\cte(\bq^2-1)}{|\bq|} +  \frac{\la  (1+\bq^2)^2  }{2\bq^2} ,
\label{wi}\\
\!\!\!\! \!\!\!\!  {\rm Hyperbolic:} &&\!\!\!\! {\cal H}_{\mathrm {MIC-Kepler}}=\frac12 (1-J_-)^2 J_+  -\frac{\cte(J_- +1)}{\sqrt{J_-}} +\frac{\la(1-J_-)^2}{2J_-}\nonumber\\[2pt]
&&\qquad\qquad \ \ =
\frac 12{(1-\>q^2)^2 }  \>p^2  - \frac{\cte(\bq^2+1)}{|\bq|} +  \frac{\la  (1-\bq^2)^2  }{2\bq^2} .
\label{wj}
\eea
It can be easily checked that in this way we have exactly recovered (in Poincaré coordinates) the curved MIC--Kepler systems studied in~\cite{Otchika, pogonerse, Otchikb}.
Obviously, all these Hamiltonians can be generalized by adding the $b_i$-centrifugal terms given in table~\ref{table2}, and in that case the QMS property is, by construction, fully preserved.


\subsection {Darboux spaces}
 
A thorough discussion of the MS potentials on the four types (I, II, IIIa and IV) of 2D Darboux surfaces was given in~\cite{KKMW03}.  Hence it is natural to compare with them the 2D versions of the  $N$D potentials given in table~\ref{table2}. To carry out this analysis, we firstly recall that the Darboux metrics given in~\cite{KKMW03} depend on two coordinates $(u,v)$, and their associated metrics and free Hamiltonians read 
 \begin{equation}
\dd s^2=F(u)^2\,(\dd u^2+\dd v^2) , \qquad
H=\frac{p_u^2+p_v^2}{F(u)^2},
\label{ds2}
\end{equation}
where $(p_u,p_v)$ are the conjugate momenta and the function $F(u)^2$ is given  by
 \be
\begin{array}{llll}
\mbox{Type I:}& \displaystyle{ F(u)^2=u } ;&\quad 
 \mbox{Type  II:}& \displaystyle{ F(u)^2=1+u^{-2} } ; \\[12pt]
 \mbox{Type IIIa:}& \displaystyle{ F(u)^2={\rm e}^{-2 u}+{\rm e}^{- u} };&\quad 
 \mbox{Type IV:}&  \displaystyle{  F(u)^2=\frac{a+\cos u}{\sin^2 u} } .
\end{array}
\ee
Secondly,   we also recall that the $N$D spaces given in table~\ref{table2} are just an $N$D spherically symmetric  generalization of these four spaces that was constructed in~\cite{PLB} through the maps 
\be
 u\to \ln r=\ln |\bq| , \qquad \dd v^2\to \dd \Omega^2 .
 \ee
 
Now, if we consider our $N$D KC and oscillator potentials given in table~\ref{table2} for these four spaces and we perform the substitution $u\equiv \ln |\bq|$, we immediately see that, in any dimension, we recover expressions for the potentials that depend {\em only} on the variable $u$. Now, if we go back to the classification given in~\cite{KKMW03}, the unique MS potentials in 2D which are functions of $u$ alone are
\be
\begin{array}{llll}
\mbox{Type I:}& \displaystyle{ \frac{1}{u}\to \frac{1}{\ln|\bq|} } ;&\quad 
 \mbox{Type  II:}& \displaystyle{ \frac{1}{1+u^2}\to \frac{1}{1+\ln^2|\bq|} } ;\\[12pt]
 \mbox{Type IIIa:}& \displaystyle{ \frac{1}{1+{\rm e}^u}\to \frac{1}{1+ |\bq|} } ;&\quad 
 \mbox{Type IV:}&  \displaystyle{ \frac{1}{a+\cos u}\to \frac{1}{a+\cos (\ln|\bq|)}} .
\end{array}
\ee

Surprisingly, we find that these four potentials are just the intrinsic {\em oscillators} of the four 2D Darboux spaces, and there is {\em no KC potential} from table 2 appearing in the classification given in~\cite{KKMW03}. This fact suggests that in the case of 2D spaces of non-constant curvature,  the intrinsic oscillator potential seems to be more fundamental than the KC one from the integrability viewpoint, since the former would be a MS system whilst the later would be only a QMS one.
 
A remark is in order: in  the classification~\cite{KKMW03}, the MS potentials associated to the Darboux spaces II and IIIa  contain also other terms depending on $u$,  namely $1/(1+u^{-2})$ and $1/(1+{\rm e}^{-u})$, respectively, which in fact  can be obtained by adding a constant $\ga$ to the above oscillator potentials with constant $\cteb$:
 \be
\begin{array}{ll}
 \mbox{Type  II:}& \displaystyle{ \frac{\cteb}{1+u^2}+\ga=\frac{\cteb+\ga}{1+u^2}+\frac{\ga}{1+u^{-2}}} ;\\[10pt]
 \mbox{Type  IIIa:}& \displaystyle{\frac{\cteb}{1+{\rm e}^u}+\ga=\frac{\cteb+\ga}{1+{\rm e}^u}+\frac{\ga}{1+{\rm e}^{-u}}}.
 \end{array}
 \label{wl}
 \ee
Finally, recall that in arbitrary dimension $N$,  only the oscillator potential with $N$ centrifugal terms  for the Darboux space IIIb has been proven to be MS~\cite{PhysD}. Therefore the MS property remains as an open problem for all the remaining types of $N$D intrinsic oscillator potentials.


\subsection { Iwai--Katayama spaces}

We have also written in table~\ref{table2}  the resulting potentials corresponding to the Iwai--Katayama spaces described in section 3.2 by considering either $a\ne 0$ or $a=0$ (with $b=1$).
We remark that, although the $\nu$-fold Kepler metric given in table~\ref{table1} depends continuously on $a$, a glance at the potentials (displayed in table~\ref{table2}) reveals that it is convenient to perform a separate analysis of the cases $a\neq0$ and $a=0$.

Firstly, let us recall that the $N$D version of the 3D multifold Kepler Hamiltonian
introduced in~\cite{IK95}  is given by
\be
{\cal H}_{\nu\mathrm {fold-Kepler}}
 =\frac { |\bq|^{2- \frac {1}{\nu}} }{2( a + b  |\bq|^{\frac 1\nu})}\left(\bp^2+
\mu^2 \bq^{-2}+ \mu^2 c\,  |\bq|^{\frac 1\nu -2}+ \mu^2 d\, |\bq|^{\frac 2\nu -2} \right) ,
\label{wm}
\ee
where   $\nu$ is a rational parameter and $a$, $b$, $c$ and $d$ are real constants. By expanding this expression we find that
\be
{\cal H}_{\nu\mathrm {fold-Kepler}}=\frac { |\bq|^{2- \frac {1}{\nu}}\bp^2 }{2( a + b  |\bq|^{\frac 1\nu})}
+\frac{\la d}{2( a |\bq|^{-\frac 1\nu} + b  ) } + \frac{\la}{2 |\bq|^{\frac 1\nu}( a + b  |\bq|^{\frac 1\nu})}
+\frac{\la c}{2( a + b  |\bq|^{\frac 1\nu})} .
\label{wn}
\ee

Hence, table~\ref{table2} allows us to provide a more clear interpretation of the four terms forming this Hamiltonian, which indeed {\em does} depend on the value of the constant $a$. Namely:

\begin{itemize}

\item If $a\ne 0$, the first term in the Hamiltonian (\ref{wn}) is the kinetic term written in table~\ref{table1}, the second is an intrinsic {\em   oscillator} with $\cteb=\la d/2$, the third is the Dirac monopole and the fourth comes out by adding a constant $\ga$  to the corresponding oscillator potential (this trick is the same as the one used in (\ref{wl})):
\be
\frac{\cteb}{a |\bq|^{-\frac 1\nu} + b }+\ga=\frac{\cteb+ b\ga}{a |\bq|^{-\frac 1\nu} + b}+\frac{a\ga}{a + b |\bq|^{\frac 1\nu} } ,
\label{wo}
\ee
so that $\ga=\la c/(2a)$. Consequently, from this approach we can state that the multifold Kepler systems with $a\ne 0$ are, in fact, {\em 
multifold oscillator systems}.  This interpretation was already given in~\cite{CQG} for the 3D case.

\item If $a=0$ (and $b=1$), the multifold Kepler Hamiltonian (\ref{wn}) reduces to
\be
{\cal H}_{\nu\mathrm {fold-Kepler};\, a=0}   =\frac 12  |\bq|^{2- \frac {2}{\nu}}\bp^2  +\frac{\la c}{2   |\bq|^{\frac 1\nu} }  + \frac{\la}{2 |\bq|^{\frac 2\nu} } +\frac{\la d}{2 } .
\label{wp}
\ee
Therefore in this case the first term is the kinetic energy given in  table~\ref{table1}, the second 
is an intrinsic {\em KC} potential with $\cte=-\la c/2$, the third is the monopole and the fourth is an additive constant. Thus only in the case $a=0$ the Hamiltonian (\ref{wn}) is a proper {\em multifold Kepler system}.

\end{itemize}

From the viewpoint adopted in this paper, it is apparent that the case $a\ne 0$ corresponds to the $\nu$-fold generalization of the 
Taub-NUT system~\cite{IK95,GW07,Ma82,AH85,GM86,FH87,GR88, BCJ, BCJM, JL}, while $a=0$ is the $\nu$-fold version of the MIC--Kepler model~\cite{IK95,KN07, NY08,mica,micb,micc}. Explicitly:

\begin{itemize}

\item 
The Taub-NUT system arises as the particular case of  (\ref{wn}) with $\nu=1$, $a=4m$, $b=1$, $c=1/(2m)$ and $d=1/(4m)^2$ (see \cite{IK95}) which yields
\bea
  && {\cal H}_{\mathrm {Taub-NUT}}=\frac { |\bq| \bp^2 }{2( 4m +   |\bq| )}
+\frac{\la |\bq|/(4m)^2}{2( 4m + |\bq|  ) } + \frac{\la}{2 |\bq| ( 4m +   |\bq| )}
+\frac{\la /(4m)}{4m +  |\bq| } \nonumber\\
&&\qquad\qquad\quad=\frac {\bp^2}{2(1+4m/|\bq|)} +\frac{\la}{2(4m)^2}\left( 1+\frac{4m}{|\bq|}\right) ,
\label{wq}
\eea
so that the same interpretation as for the multifold Kepler system with $a\ne 0$ holds. In the first line of (\ref{wq}), the first term is the kinetic energy, the second is an intrinsic {\em   oscillator} with $\cteb=\la /(2(4 m)^2)$, the third is the   monopole and the fourth corresponds to adding a constant $\ga=\la/(4m)^2$  to the oscillator potential, leading to $\cteb\to \cteb +\ga$.

\item The (flat) MIC--Kepler system is recovered from (\ref{wn}) or (\ref{wp}) when $\nu=1$, $a=d=0$, $b=1$ and $c=-2\cte/ \la$, giving rise to the expression (\ref{wh}).

\end{itemize}

Note also  that  in 3D, the generic Hamiltonian (\ref{wm}) (for any value of $a$) has been recently shown  to be  MS~\cite{CMP}. This Hamiltonian has  an additional integral of motion (coming from a generalized Laplace--Runge--Lenz vector) which, in general,  is not quadratic in the momenta.

Let us end by stressing that all the $N$D Hamiltonians that we have presented are QMS, but some of them could be MS, as we have already commented. As a matter of fact, one of the models
that we have described (namely, the Darboux III oscillator potential with monopole and centrifugal terms) is, to the best of our  knowledge,  the only known  quadratically MS system living in an
 $N$D  space of nonconstant curvature; the additional integral can be consulted in~\cite{PhysD}.
The  search for an additional integral for the rest of the systems in the Darboux spaces presented here   remains a challenging open problem. In the cases where such a constant of the motion exists, it cannot be derived from the $\rsl(2,\mathbb R)$-coalgebra symmetry, and it reflects the exceptional superintegrability properties of the Riemannian manifold determined by $f(r)$. 

On the other hand, it is worth remarking that the results presented throughout the paper can be extended as well to curved Lorentzian spaces through   an
analytic continuation method analogous to the one applied in~\cite{PLB} for some free Hamiltonians.

Finally, we would like to emphasize that the underlying $\rsl(2,\mathbb R)$-coalgebra symmetry can also be implemented in the Quantum Mechanical analogues of the systems here presented. In particular, the quantum counterparts of the integrals of the motion~\eqref{cinfm} can be readily obtained after dealing with the ordering problems which arise in the quantization of~\eqref{hamm} due to the term ${\bp^2}/{f(|\bq|)^2}$. We shall report on these and other related issues elsewhere.



\section*{Acknowledgments}

This work was partially supported by the Spanish MICINN under grants no.\    MTM2007-67389 and FIS2008-00209 (with EU-FEDER support), by CAM--Complutense University under grant  no.~CCG07-2779 (A.E.), and by the INFN--CICyT (O.R.). O.R. thanks the Einstein Foundation and Russian Foundation for
Basic Research for supporting the research project  ``Integrable/solvable
Classical and Quantal Many-Body Problems and their integrable
discretizations".  F.J.H. is  also grateful to W. Miller and A. Nersessian for very helpful
discussions.



\begin{thebibliography}{10}\frenchspacing




\bibitem{AKN97}
V.I. Arnold,   V.V.   Kozlov, A.I. Neishtadt,
 {M}athematical aspects of classical and celestial mechanics, 
Springer, Berlin, 1997.


\bibitem{BH07}
A. Ballesteros, F.J.  Herranz, 
{J. Phys. A:  Math. Theor.}
{40}  (2007) F51.   


 \bibitem{sigmaorlando}  O. Ragnisco,  A. Ballesteros,  F.J. Herranz, F.  Musso,   
 {SIGMA}  {3}   (2007)  026.  



\bibitem{GLZ92}
Y.I. Granovski, I.M. Lutzenko, A.S.  Zhedanov,
{Ann. Phys.} {217} (1992) 1. 

\bibitem{STW01}
M.B. Sheftel, P. Tempesta, P.  Winternitz, 
{J. Math. Phys.}  {42} (2001) 659.

\bibitem{Ca04}
F. Calogero,
{J. Nonlin. Math. Phys.} {11} (2004) 208. 

\bibitem{Go04}
C. Gonera,
{J. Phys. A: Math. Gen.} {37} (2004) 4085. 

\bibitem{Hu76}
A. Huaux,
{Ann. Mat. Pura Appl.} {108} (1976) 251.

\bibitem{KM84}
 E.G. Kalnins, W. Miller,
{Adv. Math.} {51} (1984) 91.

\bibitem{BMS01}
A.T. Bruce,  R.G. McLenaghan, R.G.  Smirnov,
{J. Geom. Phys.} {39} (2001) 301.




\bibitem{BHSS03}
A. Ballesteros, F.J. Herranz, M. Santander, T. Sanz-Gil,
{J. Phys. A: Math. Gen.}
{36} (2003) L93. 

 
\bibitem{vulpiPAN}
F.J. Herranz, A. Ballesteros, 
 {Phys. At. Nuclei}  {71}   (2008) 905.






\bibitem{Ev90}
 N.W. Evans,
{Phys. Rev. A} {41} (1990) 5666.


 
\bibitem{Miguel}
M.A. Rodr\1guez, P.   Winternitz, 
{J. Math. Phys.} {43}  (2002) 1309.  



 
\bibitem{Williamsx}
 E.G. Kalnins, G.C.  Williams,  W. Miller, G.S. Pogosyan,
{J. Phys. A: Math. Gen.} {35} (2002) 4755.
 



\bibitem{kiev} F.J. Herranz, A.  Ballesteros, 
   {SIGMA} {2}   (2006)  010. 


 


\bibitem{FMSUW65}
J. Fris, V. Mandrosov, Y.A. Smorodinsky, M. Uhlir, P. Winternitz,
{Phys. Lett.} {16} (1965) 354.

\bibitem{Ev90b}
N.W. Evans,
{Phys. Lett. A} {147} (1990) 483.



 \bibitem{8}
 N.W. Evans, 
{J. Math. Phys.}  {32} (1991)  3369. 


 \bibitem{10}
C. Grosche, G.S. Pogosyan, A.N.   Sissakian,
{Fortschr. Phys.} {43}  (1995) 453.

\bibitem{11}
C. Grosche, G.S. Pogosyan, A.N.   Sissakian,
{Fortschr. Phys.} {43}  (1995) 523.  




 \bibitem{20}
 E.G. Kalnins, W. Miller, G.S. Pogosyan,
{J. Math. Phys.} {38}  (1997) 5416.


\bibitem{RS}
  M.F. Rañada, M.  Santander, 
{J. Math. Phys.} {40}   (1999) 5026.




\bibitem{21}
E.G. Kalnins,  W. Miller, G.S. Pogosyan, 
{J. Phys. A: Math. Gen.} {33} (2000) 6791. 

 

\bibitem{18}
E.G. Kalnins,  J.M. Kress, G.S.  Pogosyan,   W. Miller, 
{J. Phys. A: Math. Gen.} {34} (2001) 4705. 






\bibitem{PhysD}
   A. Ballesteros, A. Enciso, F.J. Herranz, O. Ragnisco, {Physica D}   {237} (2008) 505.







\bibitem{Ca71}
F. Calogero,
{J. Math. Phys.} {12} (1971) 419.

\bibitem{Su71}
B. Sutherland,
{Phys. Rev. A} {4} (1971) 2019.

\bibitem{Mo75}
J. Moser,
{Adv. Math.} {16} (1975) 197. 

\bibitem{Wo83}
 S. Wojciechowski,
{Phys. Lett. A} {95} (1983) 279. 





\bibitem{Jauch}
 J.M. Jauch,  E.L. Hill, 
{Phys. Rev. } {57} (1940) 641. 




\bibitem{To67}
M. Toda,
{J. Phys. Soc. Japan} {22} (1967)  431. 

\bibitem{ADS06}
M.A. Agrotis, P.A. Damianou,  C. Sophocleous,
{Physica A} {365} (2006) 235. 

 \bibitem{Benentia}
   M.  Blaszak, A. Sergyeyev, 
{J. Phys. A: Math. Gen.} {38}  (2005) L1.

\bibitem{Benentib}
 A. Sergyeyev, 
{J. Math. Phys.} {48} (2007) 052114.


\bibitem{Verrier}
P.E. Verrier, N.W. Evans,
{J. Math. Phys.} {49} (2008) 022902.


\bibitem{PLB}
A.   Ballesteros, A. Enciso, F.J. Herranz, O. Ragnisco, {Phys. Lett. B}  {652}  (2007)  376.



\bibitem{BR98}
A. Ballesteros, O. Ragnisco,
{J. Phys. A: Math. Gen.}  {31} (1998) 3791.


\bibitem{Deform}
   A.  Ballesteros, F.J. Herranz,   
{J. Phys. A: Math. Gen.} {32}  (1999) 8851.


 

 \bibitem{CRMAngel}
     A. Ballesteros,  F.J. Herranz, F. Musso, O. Ragnisco, in:   
{Superintegrability in Classical and Quantum   Systems}, 
CRM Proc.~and Lecture Notes  vol.~37, 
ed.  P Tempesta {et al.}, 
 AMS, Providence, R.I., 2004, p. 1 ({\tt arXiv: math-ph/0412067}).

 



\bibitem {Doub}
B. Doubrovine,  S. Novikov,   A. Fomenko,  
{Géométrie Contemporaine, Méthodes et Applications}  First Part,
MIR, Moscow, 1982.



\bibitem{Ko72}
G. Koenigs, in:   {{L}e{\c{c}}ons sur la théorie générale des surfaces}
  vol.  4,  ed.   G. Darboux, Chelsea, New York,  1972, p. 368.



  
\bibitem{KKMW03}
E.G.  Kalnins,  J.M. Kress, W. Miller,  P. Winternitz,
 {J. Math. Phys.} {44} (2003)  5811.




\bibitem{IK95}
T. Iwai,  N. Katayama, {J. Math. Phys.} {36} (1995) 1790.




\bibitem{IK93}
T. Iwai, N. Katayama, {J. Geom. Phys.}  {12} (1993) 55.





\bibitem{EP06d}
A. Enciso, D. Peralta-Salas, {J. Geom. Phys.} {57} (2007) 1679. 

\bibitem{EP08}
A. Enciso, D. Peralta-Salas, 
Critical points and level sets in exterior boundary problems, {Indiana Univ. Math. J.}  in press.


\bibitem{LT87}
P. Li,  L.F.  Tam,   {Amer. J. Math.}  {109} (1987) 1129.


\bibitem{LT95}
P. Li,  L.F. Tam, {J. Differential Geom.} {41} (1995) 277. 




\bibitem{CQG}
 A.  Ballesteros, A. Enciso, F.J.  Herranz, O. Ragnisco, Class. Quantum Grav. {25} (2008) 165005.  


 




\bibitem{Br89}
P.J.  Braam, {J. Differential Geom.} {30} (1989) 425. 

\bibitem{SW94}
N. Seiberg, E. Witten,  {Nucl. Phys. B} {426} (1994) 19.  

\bibitem{SW94b}
N. Seiberg, E. Witten,  {Nucl. Phys. B} {431} (1994) 484.  

  

\bibitem{CK01}
S.A.  Cherkis, A. Kapustin,  {Comm. Math. Phys.}  {218} (2001)  333. 

\bibitem{KMOZ07}
P.B. Kronheimer,   T.S.  Mrowka, P. Ozsváth, 
Z. Szabó, 
{Ann. Math.}  {165} (2007) 457. 


\bibitem{KN07} S. Krivonos,  A.  Nersessian,  V. Ohanyan,    {Phys. Rev. D}  {75}   (2007)  085002.


\bibitem{MNY07} L. Mardoyan,  A. Nersessian,  A. Yeranyan,       {Phys. Lett. A}  {366}  (2007)   30.




\bibitem{NR07}
P. Norbury,  N.M. Romão,   {Comm. Math. Phys.} {270} (2007)  295. 


 \bibitem{GW07}
G.W. Gibbons, C.M. Warnick,  {J. Geom. Phys.} {57} (2007) 2286.



\bibitem{NY08}   A. Nersessian,   V. Yeghikyan,     {J. Phys. A: Math.   Theor.}  {41}  (2008)  155203.

\bibitem{Na08}
O. Nash, {Comm. Math. Phys.}  {277} (2008) 161. 










\bibitem{mica} D. Zwanziger,   {Phys. Rev.}  {176}   (1968) 1480.



\bibitem{micb} H.V. McInstosh, A. Cisneros,    {J. Math. Phys.}  {11}   (1970)  896.


\bibitem{micc} T. Iwai, Y. Uwano,    {J. Math. Phys.}  {27}   (1986) 1523.


  
 
\bibitem{Ma82}
N.S. Manton, {Phys. Lett. B} {110} (1982) 54.

\bibitem{AH85}
M.F. Atiyah,  N.J.  Hitchin, {Phys. Lett. A} {107} (1985) 21.



\bibitem{GM86}
G.W. Gibbons, N.S. Manton, {Nucl. Phys. B} {274} (1986) 183.




\bibitem{FH87}
L.G. Fehér,  P.A.  Horváthy, {Phys. Lett. B} {183}  (1987) 182.


\bibitem{GR88}
G.W. Gibbons, P.J.  Ruback, {Comm. Math. Phys.} {115} (1988) 267.


 \bibitem{BCJ} D. Bini, C. Cherubini, R.T. Jantzen,    {Class. Quantum Grav.}  {19}  (2002)  5481.

\bibitem{BCJM} D. Bini, C. Cherubini, R.T. Jantzen, B. Mashhoon,   {Class. Quantum Grav.}  {20}    (2003) 457.

\bibitem{JL} J. Jezierski, M. Lukasik,    {Class. Quantum Grav.}  {24}  (2007)  1331.




\bibitem{CMP}
 A.  Ballesteros, A. Enciso, F.J. Herranz, O. Ragnisco,
Comm. Math. Phys., to appear  ({\tt arXiv: 0810.0999}).


 
\bibitem{Otchika} V.V. Gritsev, Y.A. Kurochkin,  V.S.  Otchik,    {J. Phys. A: Math. Gen.}  {33} (2000) 4903.


\bibitem{pogonerse} A. Nersessian, G. Pogosyan,    {Phys. Rev. A}  {63} (2001) 020103.

 


\bibitem{Otchikb} A.A. Bogush, V.V. Gritsev, Y.A. Kurochkin,  V.S. Otchik,    {Phys. At. Nuclei}  {65} (2002) 1052.

 
 
 
\end{thebibliography}
\end{document}